\documentclass[a4paper]{jpconf}
\usepackage{graphicx}

\begin{document}

\title{Describe Data to get Science-Data-Ready Tooling: Awkward as a Target for Kaitai Struct YAML}

\author{Manasvi Goyal$^{1}$, Andrea Zonca$^2$, Amy Roberts$^3$, Jim Pivarski$^{1}$ and Ianna Osborne$^{1}$}

\address{$^1$ Princeton University, Princeton, NJ 08544, USA}
\address{$^2$ San Diego Supercomputer Center, La Jolla, CA 92093, USA}
\address{$^3$ University of Colorado Denver, Denver, CO 80204, USA}

\begin{abstract}
In some fields, scientific data formats differ across experiments due to specialized hardware and data acquisition systems. Researchers need to develop, document, and maintain experiment-specific analysis software to interact with these data formats. These software are often tightly coupled with a particular data format. This proliferation of custom data formats has been a prominent challenge for small to mid-scale experiments. The widespread adoption of ROOT has largely mitigated this problem for the Large Hadron Collider experiments. However, many smaller experiments continue to use custom data formats to meet specific research needs. Therefore, simplifying the process of accessing a unique data format for analysis holds immense value for scientific communities within HEP. We have added Awkward Arrays as a target language for Kaitai Struct for this purpose. 

Researchers can describe their custom data format in the Kaitai Struct YAML (KSY) language. The Kaitai Struct Compiler generates C++ code to fill the LayoutBuilder buffers using the KSY format. In a few steps, the Kaitai Struct Awkward Runtime API can convert the generated C++ code into a compiled Python module. Finally, the raw data can be passed to the module to produce Awkward Arrays. This paper introduces the Awkward Target for the Kaitai Struct Compiler and the Kaitai Struct Awkward Runtime API. It also demonstrates the conversion of a given KSY for a specific custom file format to Awkward Arrays.
\end{abstract}

\section{Introduction}
Collaborations that use a custom data format spend many hours writing their own tools to read and analyze their data. It is difficult to fund such tools because they are not usable outside the collaboration, and as a result, maintenance is difficult, leading to poor documentation and testing. It also poses a significant barrier for scientists entering the collaboration and those wishing to perform analysis outside what the original author envisioned.

Switching to a supported standard data format sounds like an obvious solution. However, in the case of the Cryogenic Dark Matter Search Collaboration \cite{cdms1-ref, cdms2-ref}, for example, this would require substantial rewriting of the data acquisition system or switching to a different one. This would require additional personnel and substantial time investment. In addition, there are legacy datasets that still have the potential for new science. This project provides a solution to this problem of reading and analyzing custom data formats for small and mid-scale collaborations across the sciences. Instead of developing their own tools, the collaborations need to describe their custom data formats in KSY format just once and then directly use the Kaitai Struct Awkward Runtime to access their data as Awkward Arrays \cite{awkward-ref}.

\section{Kaitai Struct YAML}

Kaitai-Struct YAML (KSY) \cite{kaitai-ref} is a declarative language that takes YAML-like descriptions of a data-format structure and generates parsing code in any of the supported languages. The user describes the structure of the data, not how to read or write it. KSY can describe a wide variety of file types, including image formats, network stream packet formats, and binary formats.

\subsection{An overview of Kaitai Struct}
\begin{figure}
    \centering
    \includegraphics[width=13cm]{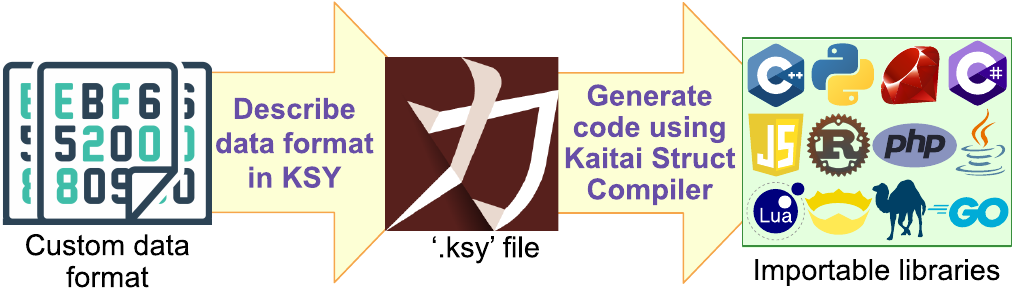}
    \caption{Kaitai Struct YAML process flowchart.}
    \vspace{12pt}
    \label{fig:kaitai_flowchart}
\end{figure}

The user describes their data structure in a `.ksy` description file using Kaitai Struct format rules \cite{kaitai-doc-ref}. The Kaitai Struct Web IDE can be used to debug the format and ensure that the data is parsed properly. The user then compiles the description file with the Kaitai Struct Compiler \cite{ksc-ref} to create importable libraries in a specified target language. These libraries include generated code for a parser, which can be included in any external project, to read the described data structure from a file or stream. The Kaitai Struct Compiler uses an extra layer of stream API to access the data to allow compilation to many target languages. This API includes Kaitai Struct runtime libraries for the respective languages used. Kaitai Struct currently supports twelve target languages including C++, Python, and Java. Figure~\ref{fig:kaitai_flowchart} shows the process of conversion of a data structure to importable libraries of the supported target languages.

\begin{figure}
    \centering
    \includegraphics[width=14.9cm]{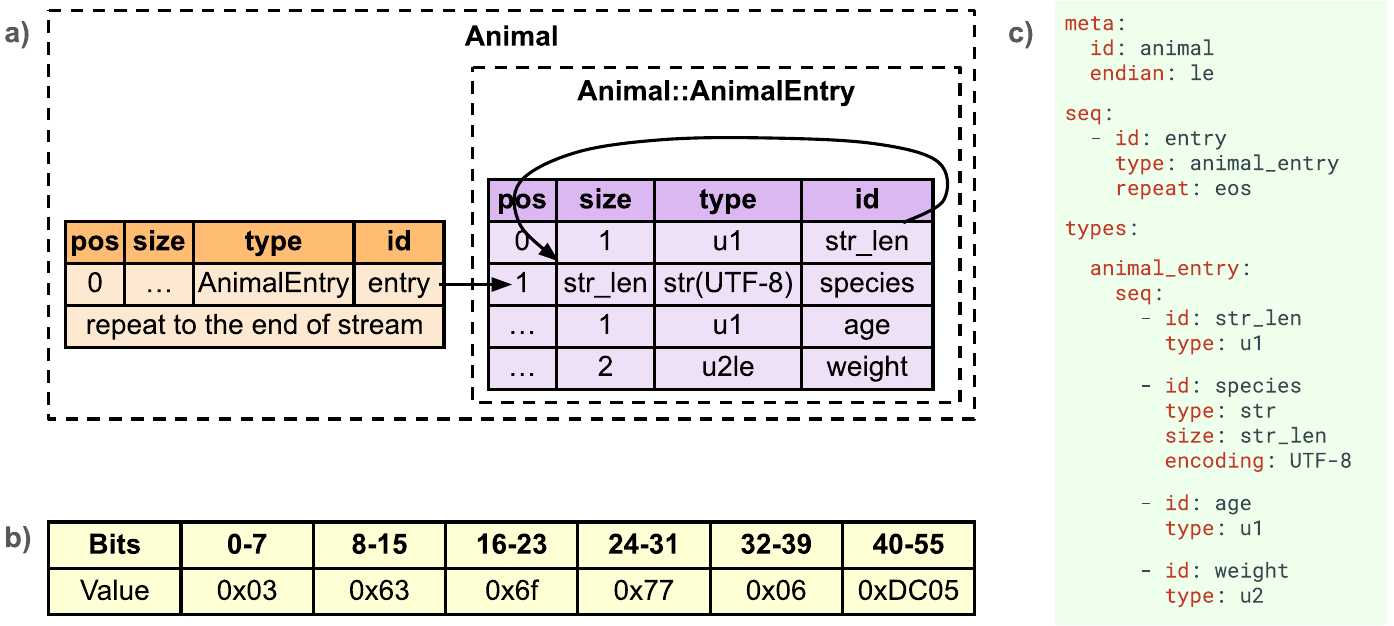}
    \caption{AnimalData Example  a) Data structure; b) An animal entry in binary; c) animal.ksy}
    \label{fig:animal_data}
\end{figure}

\subsection{Example: animalData}
Take a simple data structure, as shown in Figure~\ref{fig:animal_data}(a), containing multiple entries for animals, and each entry describes some details of the particular animal. The actual format of the binary file is slightly more complicated. Take an example with just one entry to illustrate this, as shown in Figure~\ref{fig:animal_data}(b). The first byte of this entry contains an integer value denoting the number of letters in the species name. The next N-bytes are ASCII values that represent the species name. After the ASCII characters, there is one byte for the animal's age and, finally, two bytes for the weight. Fields larger than one byte are written in little-endian format, in this case, just the weight. The given entry describes a 6-year-old cow that weighs 1500 pounds. This data structure can be described in KSY as shown in Figure~\ref{fig:animal_data}(c).

Everything in KSY starts with a \verb"meta" section. It contains the top-level information about the entire structure. The most information in \verb"meta" includes \verb"id" which specifies the name of the structure and \verb"endian" which specifies default endianness. After this section, the data structure is described using the \verb"seq" element to describe the attributes of the specific structure. Every attribute includes several keys, namely, \verb"id" which represents the attribute a name, and \verb"type" which designates the attribute type. Most real-life file formats do not contain only one copy of some element but might contain several copies, i.e., they repeat the same pattern over and over. Kaitai Struct supports the following types of repetition:
\begin{itemize}
    \item \textit{eos}: element repeated up to the very end of the stream.
    \item \textit{repeat-expr}: element repeated a pre-defined number of times.
    \item \textit{repeat-until}: element repeated while some condition is not satisfied or until it becomes true.
\end{itemize}

\section{Awkward Arrays}

The \verb"animal.ksy" is a simple example, but actual scientific data can be significantly larger with a complex structure. An example is the MIDAS event format \cite{midas-ref}, which wraps a complex but fixed binary structure with a header and footer that contain information about the length of that structure. Other scientific formats, such as the octree representations for variable-resolution data, have even more complex structures. The \verb"kaitai_struct_python_runtime" library, for example, stores all data structures as a dictionary. While this is a familiar interface to many users, it is also extremely slow for files even in the MB range.

Awkward arrays, a Scikit-HEP \cite{scikithep-ref} library, offer a powerfully dynamic and efficient approach to represent complex data structures in Numpy-like arrays \cite{numpy-ref, awkward-scipy2020-ref}. They store data in jagged nested arrays of arbitrary types and variable lengths while maintaining speed \cite{awkward-paper-ref}. These capabilities make Awkward an appealing target language for Kaitai Struct. Therefore, the C++ header-only \verb"LayoutBuilder" \cite{layoutbuilder-ref} libraries of Awkward Array are used to represent the scientific data formats in Numpy-like arrays.

\section{Awkward Target for KSY}

The objective is to combine the features of Awkward with the Kaitai Struct ecosystem. Awkward target has been added to \verb"kaitai_struct_compiler" as an \verb"AwkwardCompiler" to adapt the existing \verb"CppCompiler" to build Awkward arrays. The user only needs to focus on describing their particular data structure into \verb"'.ksy'" format. They have to do this only once for a specific format. The entire code generation is handled by the \verb"kaitai_struct_awkward_runtime" \cite{ksar-ref}. With a few simple commands, the user can load their raw data file in the Python module to represent their data in well-structured Awkward Arrays.

\begin{figure}
    \centering
    \includegraphics[width=14.5cm]{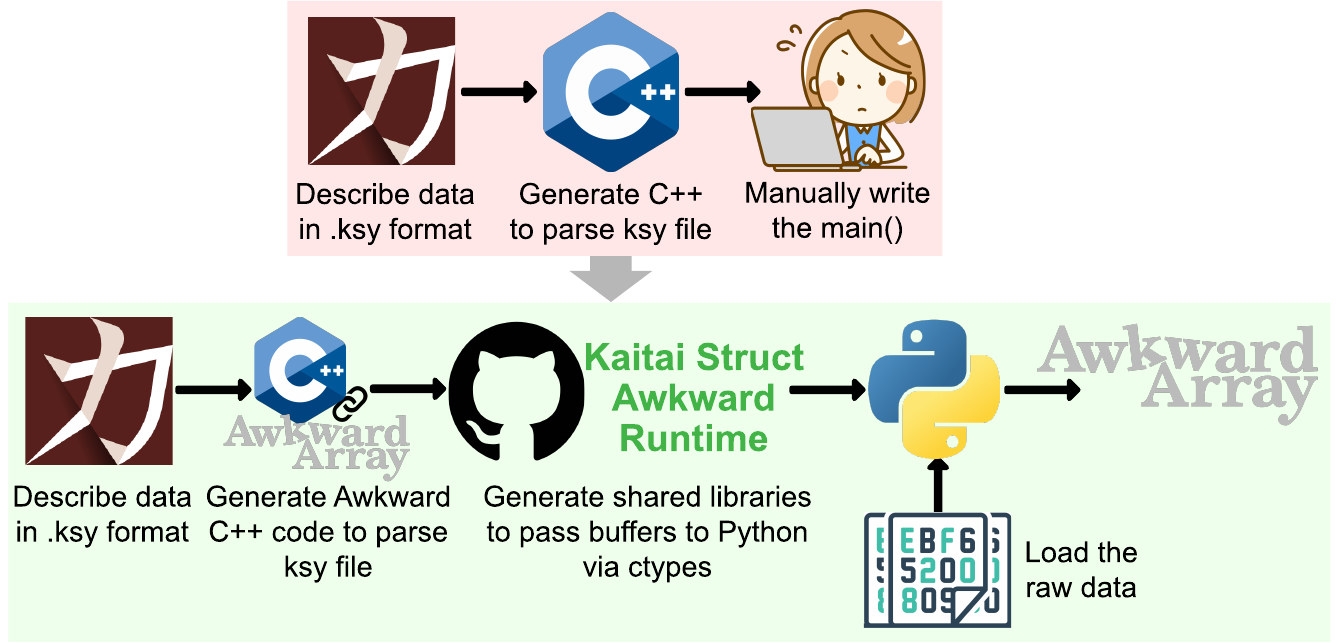}
    \caption{Kaitai Struct C++ Runtime vs Kaitai Struct Awkward Runtime}
    \label{fig:bad_vs_good}
\end{figure}

\subsection{Kaitai Struct Awkward Runtime User Interface}
As shown in Figure~\ref{fig:bad_vs_good}, the user always starts by describing their data structure in KSY format. Then, use \verb"kaitai_struct_compiler" with the desired target language to generate the modules that parse the KSY file. When users generate code with C++ target, they must write a \verb"main()" function to access the data, which can be time consuming when dealing with complex scientific data. 
However, when the user generates the parsing code with the Awkward target, containing \verb"LayoutBuilder" filling instructions \cite{awkward-kaitai-pyhep2023}, they can load the generated shared library (.so) and the raw data into Awkward Arrays in a Python environment and can immediately start writing their analysis code.

\begin{figure}
    \centering
    \includegraphics[width=13.7cm]{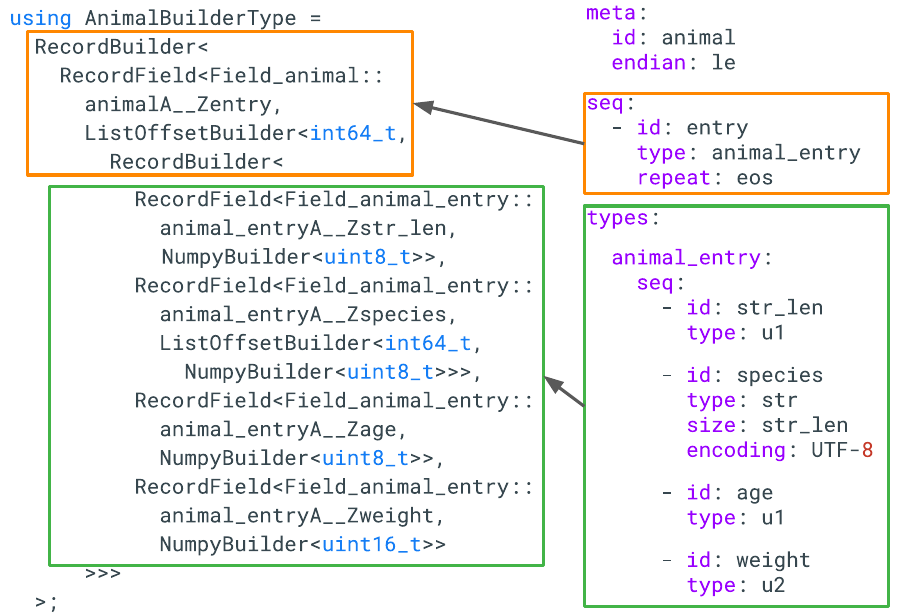}
    \vspace{-2pt}
    \caption{KSY data structure as LayoutBuilder.}
    \label{fig:ksy_layoutbuilder}
\end{figure}

\subsection{Procedure to use Kaitai Struct Awkward Runtime}

The \verb"kaitai_struct_awkward_runtime" repository contains the step-by-step procedure \cite{ksar-ref} to use the Kaitai Struct Awkward Runtime to generate Awkward arrays for a given KSY file.
Once the user has installed the required dependencies, they can use the \verb"kaitai-struct-compilper" tool to generate the parsing code for \verb"awkward" target. Figure 4 shows the representation of the KSY structure in LayoutBuilder code using the example in Section 2.2. The generated LayoutBuilder code can be compiled into a shared library using a \verb"awkward-kaitai-build" command with the required command line options.

The procedure involves the installation of the \verb"awkward-kaitai" module using \verb"pip" to enable the integration of LayoutBuilder C++ code with Python via ctypes allowing dynamic linking and execution through the generated shared library. The \verb"awkward-kaitai" package provides the necessary Python tools to connect the data structures of the parsing code with Awkward Array operations. 

Now, the user can open the Python terminal and import the \verb"awkward_kaitai" module. They can pass the path of the generated static library to the \verb"Reader" class and then pass the path of the raw data file to the \verb"load" function to fill the Awkward array layout with the raw data. Finally, the generated \verb"ak.Array" can be printed or used for further analysis. An example of the Python terminal steps using the \verb"animal.ksy" is as follows.

\begin{verbatim}
import awkward_kaitai
animal = awkward_kaitai.Reader("./src-animal/libanimal.so")
awkward_array = animal.load("example_data/data/animal.raw")
awkward_array.to_list()
\end{verbatim}

\noindent Finally, \verb"animal.ksy" is represented in Awkward Arrays as:

\begin{verbatim}
[{'animalA__Zentry':[
    {'animal_entryA__Zstr_len': 3, 'animal_entryA__Zspecies': 'cat',
    'animal_entryA__Zage': 5, 'animal_entryA__Zweight': 12},
    {'animal_entryA__Zstr_len': 3, 'animal_entryA__Zspecies': 'dog',
    'animal_entryA__Zage': 3, 'animal_entryA__Zweight': 43},
    {'animal_entryA__Zstr_len': 6,'animal_entryA__Zspecies': 'turtle',
    'animal_entryA__Zage': 10, 'animal_entryA__Zweight': 5}
]}]
\end{verbatim}

\section{Conclusion}
Awkward Target for Kaitai Struct provides a promising solution for custom scientific data formats in small to mid-scale experiments. It leverages the descriptive power and ecosystem of the Kaitai Struct and combines it with the features of Awkward Arrays. Researchers can easily convert the KSY descriptions of their custom data formats into analysis-ready code using the Kaitai Struct Awkward Runtime API. This approach holds the potential to benefit various scientific disciplines, within and beyond HEP.

\section{Acknowledgment}
This work is supported by NSF cooperative agreements OAC-1836650 and PHY-2323298 (IRIS-HEP) and grants OAC-2104003 (PONDD) and OAC-2103945 (Awkward Array).

\end{document}